\documentclass[twocolumn]{article}

\usepackage[utf8]{inputenc}
\usepackage{graphicx}         
\usepackage{amsmath}          
\usepackage{url}              
\usepackage{booktabs}         
\usepackage{hyperref}         
\usepackage[a4paper, total={7in, 9.5in}]{geometry} 
\usepackage{caption}          
\usepackage{array}            

\hypersetup{
    colorlinks=true,
    linkcolor=blue,
    filecolor=magenta,      
    urlcolor=cyan,
    citecolor=blue,
}

\title{\textbf{A Knowledge Graph-based Retrieval-Augmented Generation Framework for Algorithm Selection in the Facility Layout Problem}}

\author{
    Nikhil N S\textsuperscript{1},  Bilal Muhammed\textsuperscript{2}, Soban Babu Beemaraj\textsuperscript{2}, Amol Joshi\textsuperscript{2}\\[1ex] 
    \textsuperscript{1}Indian Institute of Science, Bengaluru, India\\
    \texttt{nikhilns@iisc.ac.in}\\[2ex] 
    \textsuperscript{2}TCS Research and Innovation, Tata Consultancy Services Ltd.\\
    \texttt{
    bilal.muhammed@tcs.com,
    soban.bb@tcs.com, amol.joshi@tcs.com}
}

\date{} 

\begin{document}

\maketitle

\begin{abstract}
\noindent Selecting a solution algorithm for the Facility Layout Problem (FLP), an NP-hard optimization problem with multiobjective trade-off, is a complex task that requires deep expert knowledge. The performance of a given algorithm depends on the specific characteristics of the problem, such as the number of facilities, objectives, and constraints. This creates a need for a data-driven recommendation method to guide algorithm selection in automated design systems. This paper introduces a new recommendation method to make this expertise accessible, based on a Knowledge Graph-Based Retrieval-Augmented Generation (KG-RAG) framework. In this framework, a domain-specific knowledge graph (KG) is constructed from the literature. The method then employs a multifaceted retrieval mechanism to gather relevant evidence from this KG using three distinct approaches: precise graph-based search, flexible vector-based search, and cluster-based high-level search. The retrieved evidence is utilized by a Large Language Model (LLM) to generate algorithm recommendations based on data-driven reasoning. This KG-RAG framework is tested on a use case consisting of six problems comprising of complex multi-objective and multi-constraint FLP case. The results are compared with the Gemini 1.5 Flash chatbot. The results show that KG-RAG achieves an average reasoning score of 4.7 out of 5 compared to 3.3 for the baseline chatbot. 
\end{abstract}

\noindent\textbf{Keywords:} Facility Layout Problem, Algorithm Selection, Knowledge Graph, Retrieval-Augmented Generation, Large Language Models, Recommender systems.

\section{Introduction}
\label{sec:introduction}
The Facility Layout Problem (FLP) is an important part of industrial engineering and operations research \cite{ref1}. Because the operating cost and service performance are directly influenced by effective layout design. The suitable placement of facilities reduces the total operating costs by 20\% to 50\% \cite{ref2}. The computational complexity of FLP stems from its combined discrete and continuous nature. In practice, this complexity is magnified by multi-objective trade-offs. For example, minimize material handling costs and maximize safety. This multi-objective problem is coupled with constraints such as aspect-ratio, adjacency and circulation constraints, and dynamic considerations such as reconfigurations to accommodate product-mix changes, demand changes, or equipment changes. These considerations render many FLP variants NP-hard, so exact mathematical programming is computationally tractable only for small or specially structured instances \cite{ref3}. This FLP is also dynamic in nature because of evolving real-world demand. Selecting the algorithm for the FLP is challenging and requires expert knowledge because of the complex nature of the FLP.

The solution techniques have been evolving as the complexity of the FLP increases, as mentioned above. Initial solutions, such as quadratic assignment problem \cite{ref4}, mixed-integer programming, and graph-theoretic formulations \cite{ref5} for equal-area FLP. These approaches guarantee optimal solutions; however, the exponential growth in computational time severely limits their practical use in industrial-scale problems. As the problem size grows, researchers shifted to heuristic approaches, such as construction and improvement heuristics. Here, the construction heuristics build layouts from the ground up, for example, an automated layout design program and computerized relationship layout planning \cite{ref6}, whereas the improvement heuristics build solutions incrementally, for example computerized relative allocation of facilities technique \cite{ref6}. These techniques are likely to converge to suboptimal layouts. To enhance the ability of handling large and complex search spaces, various metaheuristic techniques, including particle swarm optimization, genetic algorithms, ant simulated annealing, colony optimization, and others, have been introduced \cite{ref7}. These techniques increase the likelihood of avoiding local optima. Moreover, the hybrid metaheuristic approaches, such as biased random-key genetic algorithms with linear programming repair, have recently been developed as state-of-the-art solutions, with enhanced solution quality and scalability \cite{ref8}. 

Despite these developments, the situation in FLP today is a broad and scattered collection of niche algorithms. This is typical of the no free lunch theorems, which state that when performance is averaged across all possible optimization problems, no single algorithm is universally superior. For example, an algorithm that performs well in small-scale, static settings may be desired in large-scale, multi-objective, or dynamic conditions \cite{ref9}. As a result, attention has turned to the meta-problem of algorithm selection, which involves selecting the algorithm, representation, and parameters best suited to an instance of information \cite{ref21}.

The early attempts to address this meta-problem were static decision-support methods. For instance, a GenOpt-based \cite{ref10} framework was developed to guide practitioners in selecting from a wide range of general optimization algorithms. Their approach, which utilized flowcharts and choice matrices, was designed to assist novices by matching algorithm characteristics to problem types and performance requirements. While this represents a valuable step towards systematizing algorithm selection, the framework was not specific to the unique complexities of the FLP. Furthermore, like other static methods, it is not dynamic and cannot adapt to new algorithms or shifting problem definitions without manual updates \cite{ref10}. To improve flexibility, research employed data-driven approaches, namely meta-learning, that predict algorithm performance from problem meta-features, for example, size, constraints, and statistical properties \cite{ref11}. SATzilla, for example, selects the fastest solver from a portfolio based on instance features and has demonstrated strong performance in combinatorial optimization \cite{ref12}. This approach later evolved into AutoML, where algorithm selection is combined with hyperparameter tuning under the Combined Algorithm Selection and Hyperparameter (CASH) framework \cite{ref13}. Although adaptive frameworks and AutoML have built significant capabilities, the emergence of Large Language Models (LLMs) such as OpenAI’s GPT series and Google’s Gemini introduces new possibilities \cite{ref14}. Trained on large, general-purpose corpora, these models are well-suited for acquiring language, reasoning, and in-context learning; however, when applied to domain-specific problems such as FLP optimization they can produce hallucinations (plausible but incorrect responses) a risk that is exacerbated by their weak domain adaptation  \cite{ref15} \cite{ref16}.

AI research has increasingly focused on grounding LLMs with structured and verifiable knowledge to reduce hallucinations and improve domain adaptation. Within this direction, two complementary paradigms have gained prominence: Retrieval-Augmented Generation (RAG) and Knowledge Graphs (KG). RAG retrieves relevant domain-specific information before generation, mitigating hallucinations and enabling integration of up-to-date content without retraining \cite{ref17}. KG organise domain entities and relationships, enabling explainable reasoning and mitigating data sparsity or cold-start issues. The relational pathways in KG enhance accuracy and interpretability, which are critical for decision support in technical domains \cite{ref18}. A new approach integrates these paradigms. A Knowledge Graph-Based Retrieval-Augmented Generation  (KG-RAG) \cite{ref19} model uses a structured KG as the retrieval source for an RAG. The KG offers an explainable, linked domain model, and RAG allows LLMs to flexibly query and blend this knowledge into transparent and evidence-based proposals. This integration is suitable for comprehensive domain knowledge and justifiable output for complex recommendation tasks. The literature indicates a convergence of a long-standing operations research problem with a contemporary AI opportunity. Algorithmic fragmentation in FLP, as implied by the no free lunch principle, makes the task of algorithm selection a critical bottleneck. Current machine learning-based selectors, though adaptive, tend to operate as black boxes. Concurrently, the KG-RAG paradigm addresses Large Language Model (LLM) limitations of hallucinations and poor performance in domain adaptation while allowing for structured and transparent reasoning.

This paper introduces a dynamic and domain-specific solution using the KG-RAG framework to generate data-driven algorithm recommendations for the FLP, given a user-defined query and a domain-specific KG of historical solutions. Our approach begins with constructing a formal KG that captures the complex relationships between FLP instances, their unique characteristics, and the performance of various solution methodologies. To leverage this structured knowledge base, the method employs a multi-faceted retrieval pipeline that integrates precise graph queries, flexible vector search, and high-level cluster analysis to gather comprehensive evidence. An LLM then synthesizes this evidence to generate data-driven and explainable recommendations for a complete methodology, including the algorithm, algorithm parameters, a suitable problem representation, and a constraint handling technique. This work includes an intelligent feedback loop, which allows the method to function as a continuously learning environment by incorporating new user-submitted solutions, thus ensuring its long-term relevance and accuracy.

The paper has been structured as follows: Section 2 presents the proposed architecture and methodology, detailing the construction, domain-specific KG, and the multi-faceted retrieval and recommendation pipeline design. Section 3 provides a comparative experimental evaluation, benchmarking the KG-RAG method against the Gemini 1.5 Flash LLM chatbot to empirically validate its performance on recommendation accuracy and its ability to generate data-driven, interpretable reasoning. 

\section{Architecture and methodology}
\label{sec:architecture}
The FLP concerns placing a set of facilities (machines, departments, storage areas, etc.) inside a bounded site to optimize one or more performance measures.  While Figure 1 provides a simple illustration of the FLP layout context, the practical problem spans multiple dimensions that materially affect solver choice and complexity. These begin with geometric representation, from the equal-area rectangles of early models \cite{ref4} to the unequal area, aspect-ratio rectangles and general polygons of modern applications \cite{ref1}. The problem scale is equally decisive \cite{ref18}, with the objective structure further constraining algorithm suitability, as single-objective methods, for example, minimizing material-handling cost, efficient area utilisation, minimizing material flow, may be ineffective for multi-objective trade-offs, for example, cost vs. safety vs. maintainability and vice versa \cite{ref20}. Then the constraint coupling fundamentally shapes the feasibility landscape, including non-overlap, adjacency requirements, boundary constraints, and other geometric or operational limits. Together, these factors create a vast search space, leading to high variability in algorithm performance across different FLP instances. This diversity, as discussed in Section 1, underscores the necessity for a specialised algorithm recommendation method to adjust to each case's unique features. 
\begin{figure}[htbp]
    \centering
    \includegraphics[width=0.9\columnwidth]{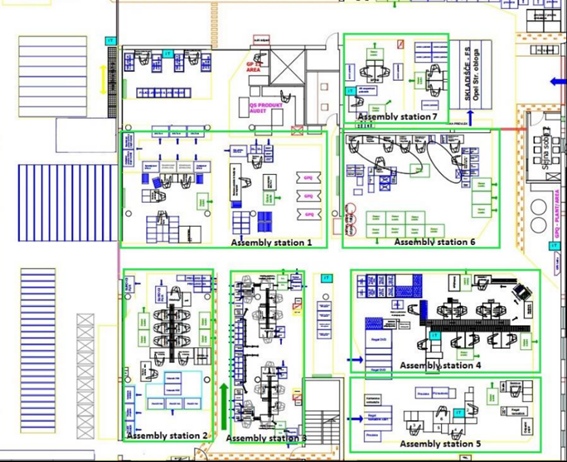}
    \caption{Example facility layout of production area [28]}
    \label{fig:layout_example}
\end{figure}

Bridging this gap requires a systematic algorithm-selection approach that moves beyond generic recommendations. An ideal method must perform instance-aware matching, aligning suggestions with the specific scale, geometry, and constraints of the input account for objective type, ensuring alignment with single- or multi-objective goals and providing constraint-handling guidance, recommending methods, for example, repair operators, penalty functions, or exact feasibility checks. Beyond naming a method, it should supply validated algorithm parameters, such as population size for a genetic algorithm and starting temperature and cooling schedule for simulated annealing, as users often lack expertise in setting effective hyperparameters. For adoption, the method must ensure explainability and provenance, justifying recommendations with historical performance data, and deliver actionable outputs.

\subsection{Architectural overview}
\begin{figure}[htbp]
    \centering
    \includegraphics[width=1\columnwidth]{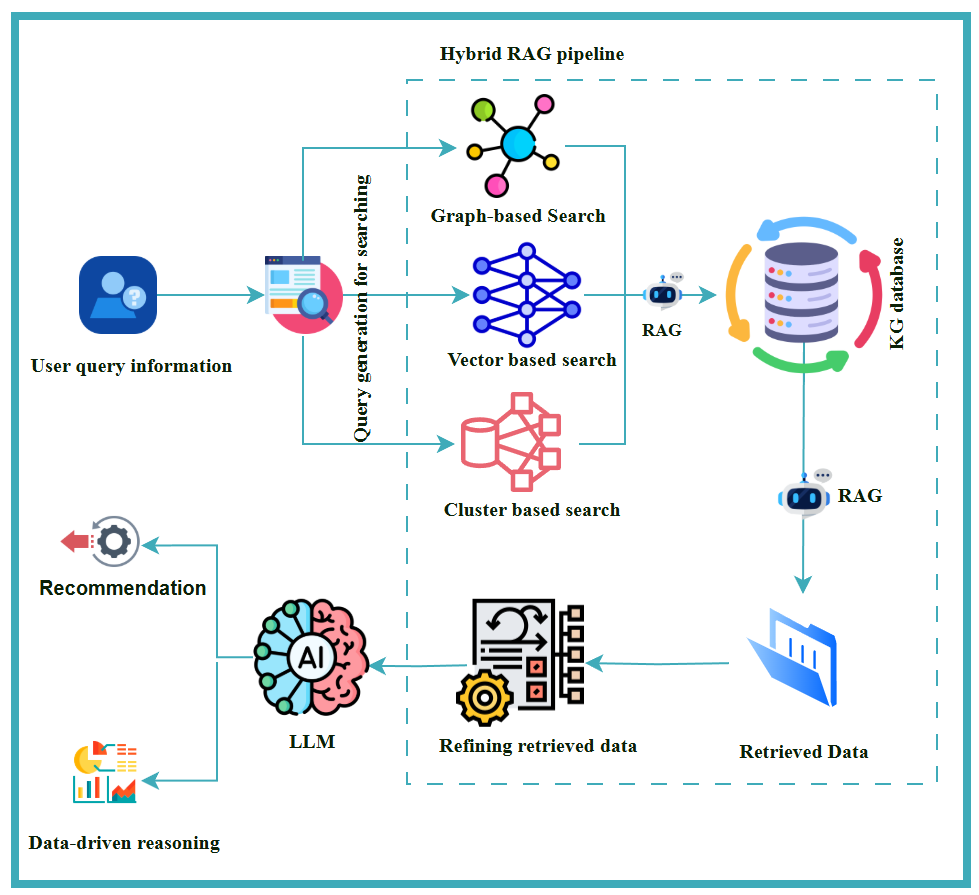}
    \caption{High-level workflow of the recommendation}
    \label{fig:method_overview}
\end{figure}

A recommendation method is proposed to address the previously identified requirements. The method's architecture, depicted in  Figure~\ref{fig:method_overview}, comprises several key components, each designed to address a specific research gap, as explained below.

\begin{figure*}[htbp]
    \centering
    \includegraphics[width=1\textwidth]{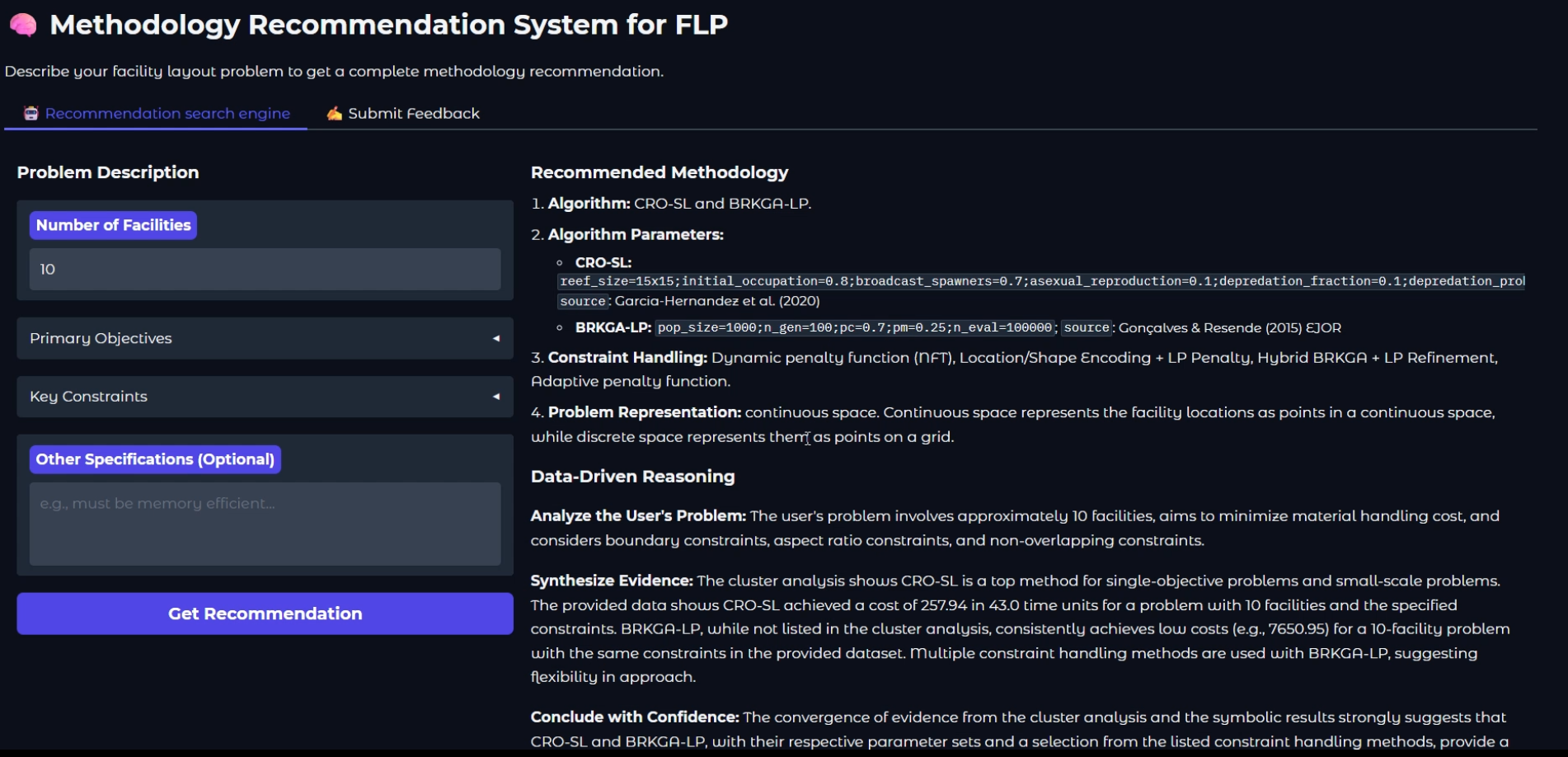}
    \caption{User Interface}
    \label{fig:ui}
\end{figure*}

\begin{itemize}
    \item \textbf{Initialisation of user query input:} The process begins with the user defining their FLP. This component addresses the need for input flexibility, allowing users to specify diverse combinations of objectives, the number of facilities, and constraints, moving beyond the rigid definitions. 
    \item \textbf{Hybrid evidence retrieval:} The method initiates a parallel, three-streamed retrieval process to overcome the limitations of static and single-faceted search methods. This component provides a comprehensive and adaptable search by collecting evidence from the \href{https://neo4j.com/}{Neo4j} KG. This includes:
    \begin{itemize}
        \item \textbf{Graph-based search:} A precise graph-based search using Cypher to find exact or structurally similar problems. 
        \item \textbf{Vector-based search:} A dynamic vector-based search to find conceptually related problems based on the user's text description.
        \item \textbf{Cluster-based search:} An aggregate analysis to identify high-level trends for the given problem type, for example, top algorithms for large-scale, multi-objective problems.
    \end{itemize}
    \item \textbf{Evidence compilation:} The results from all three retrieval channels are integrated and compiled into one contextualized dossier. The evidence is then organized into a close-fitting prompt template with the user's original query to anchor the language model.
    \item \textbf{LLM-powered recommendation:} To address the black box nature and the risk of LLM hallucinations, the evidence from all three retrieval channels is passed to this component. The results are compiled into a contextualized dossier that grounds the Google Gemini LLM. The LLM is prompted not to use its general knowledge but to act as an expert analyst, compiling the provided data into a final, formatted recommendation with a clear, evidence-driven explanation.
    \item \textbf{Continuous learning via feedback:} The design has a feedback loop. When users add new solved problem cases, the method learns and incorporates this information smartly, enriching the KG.

\end{itemize}
The method is implemented in Python, utilizing a modern technology stack composed of Neo4j for the graph database and vector index, Google Gemini as the LLM and embedding engine, LangChain for orchestrating the RAG pipeline, and Gradio for the interactive web user interface. The user interacts with the system through the interface shown in Figure~\ref{fig:ui}. They can input specific problem parameters, such as the number of facilities and key constraints, to receive a comprehensive methodology recommendation, including the selected algorithm, its parameters, and the underlying data-driven reasoning.
\subsection{Knowledge graph construction}
The method's foundation is a designed KG that transforms a flat, tabular dataset into a rich, interconnected network of domain knowledge.
\begin{table}[htbp]
\caption{Schema of a CSV source file}
\label{tab:csv_schema}
\centering
\resizebox{\columnwidth}{!}{%
\begin{tabular}{lll}
\toprule
Column
& Description
& \textbf{Type} 
\\
\midrule
Problem ID
& Problem identifier
& string 
\\
Number of Facilities& Number of Facilities
& integer 
\\
Floor Dimensions
& Width × Height (m)
& float 
\\
Problem Representation
& Layout type
& string 
\\
Facility Data& Dimensions type
& string 
\\
Objective
& Optimization goal
& string 
\\
Constraints
& Layout restrictions
& string 
\\
Constraint Handling& How constraints are handled
& string 
\\
Optimization Method
& Solver used
& string 
\\
Model Parameters
& Key algorithm parameters
& String 
\\
Solution Cost
& Solution cost
& float 
\\
Computation Time
& Computation time
& float 
\\
Source
& Reference
& String \\
\bottomrule
\end{tabular}
}
\end{table}

\subsubsection{Data corpus and schema}
The initial knowledge base was derived from a CSV file, a comprehensive dataset aggregating solved FLP instances from academic literature and benchmarks. Table~\ref{tab:csv_schema} summarizes the structure of this dataset, where each row represents a layout problem and its solution. The columns capture key information such as the problem ID, number of facilities, floor dimensions, problem type, constraints, solver method, algorithm parameters, solution cost, computation time, and source reference.

\subsubsection{Graph schema design}
This tabular data, as Table~\ref {tab:csv_schema} schema, is loaded into a Neo4j graph database through a custom Cypher script. The graph schema, used in Table~\ref{tab:node_schema}, and ~\ref{tab:relationship_schema} is defined to describe the domain entities and their complex relationships explicitly\textbf{.}

\begin{table}[htbp]
\caption{Nodes and properties schema}
\label{tab:node_schema}
\centering
\begin{tabular}{>{\raggedright\arraybackslash}p{0.45\linewidth}l}
\toprule
\textbf{Node label}& \textbf{Key properties}\\
\midrule
Problem & problem\_id, num\_facilities\\
Method & name \\
Objective & name \\
Constraint & name \\
Representation & name \\
ConstraintHandling & name \\
Solution & id (unique), cost, time\_sec\\
ObjectiveCluster & classification of  objective\\
ScaleCluster & Classification on problem's size\\
ConstraintHandlingCluster & category of cons\_ handling\\
\bottomrule
\end{tabular}
\end{table}
\begin{table}[htbp]
\caption{Relationship schema}
\label{tab:relationship_schema}
\centering
\begin{tabular}{l}
\toprule
\textbf{Relationship structure}\\
\midrule
(Solution) - [solved] $\rightarrow$ (Problem)\\
(Solution) - [used\_method] $\rightarrow$ (Method)\\
(Problem) - [has\_objective] $\rightarrow$ (Objective) \\
(Problem) - [has\_constraint] $\rightarrow$ (Constraint) \\
(Problem) - [has\_representation] $\rightarrow$ (Representation) \\
(Problem) - [cons\_handling] $\rightarrow$ (ConsHandling)\\
(Problem) - [belongs\_to\_scale] $\rightarrow$ (ScaleCluster) \\
(Method) - [is\_type\_of] $\rightarrow$ (MethodCluster) \\
(ConsHandling) - [is\_type\_of] $\rightarrow$ (ConsHandl
Cluster)\\
(Problem) - [objective\_cluster] $\rightarrow$ (ObjectiveCluster)\\
\bottomrule
\end{tabular}
\end{table}

\subsubsection{Automated clustering and vector-based indexing}
The KG is enriched with categorical clusters and vector-based embeddings for higher-level reasoning.
\begin{itemize}
    \item \textbf{Automated clustering:} Following the first import of data, a sequence of Cypher queries automatically clusters nodes according to deterministic rules. For instance, issues are clustered into a scale cluster by their number of facilities and into an objective cluster by whether the objective name contains a comma, which is a reasonable heuristic for multi-objective problems. This rule-based, automated mechanism brings consistency and scalability with the later additions of data appended.
    \item \textbf{Embedding generation:} Each problem node is augmented with a vector embedding to enable conceptual search. Key attributes, such as objective and constraints, induce a full-text description property. The text is passed through the Google AI embedding model, and the resultant vector is cached in the node and indexed into Neo4j's native vector index. The process is done once on new data, indexing all problems for similarity search.
\end{itemize}

\subsection{The hybrid RAG recommendation pipeline}
Whenever the user provides the query (number of facilities, objectives, constraints, and other specifications, if any), the method initiates a multi-stage pipeline to gather complete evidence before making a recommendation. After submission, the method normalizes and validates the inputs. The selections from the dropdowns are treated as validated entities. The free-text entries are treated as unvalidated entities. This distinction is critical as it dictates how the graph-based search query is constructed.

\textbf{Adaptive query generation:} The algorithm generates a set of evidence-gathering queries from the validated and unvalidated inputs. The initial step is exhaustive and serves as the basis for the rest of the steps of the RAG pipeline. Also, the population embedding process happens at method boot time and, significantly, is re-run for every new instance of the problem added by the user feedback loop.

\subsubsection{Multi-faceted evidence retrieval}
The method employs three parallel retrieval strategies to build a rich context for the LLM.
\begin{itemize}
    \item \textbf{Graph-based search:}  The algorithm performs an exact, structured search in the KG. A Cypher query is dynamically built from user inputs. The query has a package of intelligent features loaded for exceptional cases.
    \begin{itemize}
        \item \textbf{Flexible filtering:} The query applies an or constraint between constraints and goals to prevent the search from narrowing down too much. Besides that, it also dynamically regulates the quantity of facilities. Instead of searching for an exact amount, the Cypher query searches for problems in a ± 25\% range of the number of facilities given by the user. This dynamic range significantly increases the possibility of retrieving meaningful historical information, even if no exact number matches the facility number in the KG. 
        \item \textbf{Performance-oriented ordering:} The retrieved results are not random; a multi-factor scoring method ranks them to favour the most holistically suitable solutions. This relevance-first prioritises finding the correct solution over merely finding a high-performing one. The ORDER BY clause now prioritises results in a high-level order: first by the highest objective score (best matching the most user-specified objectives), second by the highest constraint score (best matching the most user-specified constraints). Only as a final resort does it fall back to using proximity to the user's number of facilities, the minimum cost, and the minimum time as tiebreakers. This means that the highest-ranked results are high-performing and, most importantly, relevant to the user's problem. 
        \item \textbf{Multi-objective handling}: If the user picks more than one objective, the query logic is then optimized to favour pulling solutions for issues that are tagged explicitly in the multi-objective category within the KG. This ensures that pulled examples are relevant to the problem of balancing competing objectives. 
        \item \textbf{Adaptive fallback for unprecedented scale: }The method fails gracefully when a user query is beyond the range of the current knowledge base. During boot time, the method preloads the maximum number of facilities in the KG. When its initial graph-based search fails and the requested number of facilities by the user is greater than this preloaded maximum, a fallback query is triggered programmatically. The query smartly broadens its search to find the most relevant large-scale precedents. It retrieves solutions from problems that meet two conditions. Problems with a facility count in the top quartile of the dataset's known scale. Problems belonging to the top-level scale cluster. 
    \end{itemize}

    \item \textbf{Vector-based search:} The user's free-text problem description is translated into a vector representation. The vector then performs a similarity search against the indexed problem nodes in Neo4j. This returns conceptually similar problems. 
    \item \textbf{Cluster-based search:} To further support the provision of high-level statistical context, the method analyses the list of problem IDs found by the graph-based search (before applying the final limit). It then aggregates queries to find the top three most common patterns in solutions for the scale cluster and objective cluster associated with the user query. This provides both strong trend-based evidence and instance-level results.
\end{itemize}

\subsection{LLM compilation and generation}
All the top-ranked graph-based outcomes, top vector-based matches, and cluster analysis trends retrieved evidence are combined into one comprehensive context. The dossier is written in a strict prompt template, with the original user query submitted to the Google Gemini LLM. The prompt tells the model to write as a professional analyst, combining all the presented evidence to answer the form of two sections: a straightforward recommendation and a clear data-driven reasoning part, clearly explaining the recommendation using the retrieved data. Such an approach is required to guarantee the output's trust and transparency. By tightly coupling the LLM with the provided context, the method architecturally rules out model hallucination risk and refrains from making recommendations on untested assumptions.

Finally, it incorporates a continuous learning loop to ensure the method remains current and avoids the static knowledge base problem. When users provide new solved problem instances, a backend process ensures stable and autonomous knowledge integration. This includes data standardization, such as combining multiple objectives into a single, canonical string to maintain uniformity in the KG. The method then utilizes Cypher's MERGE command to intelligently update the graph, either linking to existing entities or creating new ones without duplication. This feature enables dynamic and continuous learning, enriching the knowledge base and allowing the method to make more accurate and relevant recommendations in the future.

\subsection{Comparative evaluation}
This section presents the experimental design and results of a comparative study conducted to evaluate the performance of our proposed KG-RAG method. The primary objective is empirically confirming that grounding an LLM in a well-structured, domain-specific KG leads to significantly accurate and interpretable recommendations than a baseline LLM approach using raw, structured data.

\subsubsection{Evaluation methodology}
The evaluation is designed as a direct comparative analysis between our proposed model and a strong baseline model, with five distinct test cases designed to cover a range of real-world FLP scenarios, from simple, single-objective problems to more complex multi-objective and ill-defined queries. 

\begin{itemize}
    \item \textbf{Baseline model:} The baseline Google Gemini 1.5 Flash LLM as the proposed model, but instead of a KG, it is provided with the entire CSV dataset file as its knowledge source. This represents a powerful, traditional RAG method that relies on the LLM's ability to find and reason with information from a flat, tabular text file. 
    \item \textbf{Ground truth establishment:} For each test case, a ground truth recommendation was established to provide a fair benchmark for accuracy. This was not a subjective choice but the outcome of a systematic manual analysis of the source dataset. Figure~\ref{fig: ground_truth}  illustrates the process, which mirrors the logic of ground truth determination.  
\begin{figure}[htbp]
    \centering
    \includegraphics[width=0.9\columnwidth]{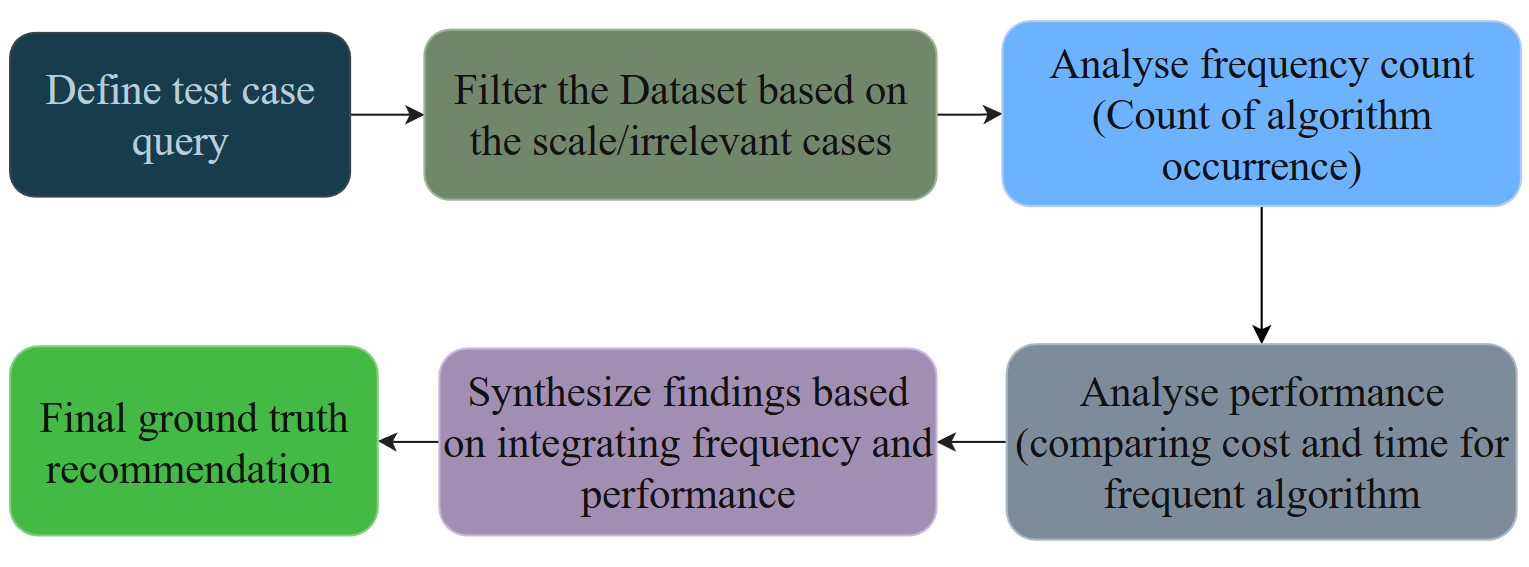}
    \caption{Ground truth establishment flow }
    \label{fig: ground_truth}
\end{figure}

    \textbf{{Metrics:}} Each method's performance was measured against the ground truth using two key criteria.
    \begin{enumerate}
    \item \textbf{Recommendation accuracy:} binary measure (1 for a match, 0 for a mismatch) of whether the method's top recommended algorithm(s) matched or included the ground truth. 
    \item \textbf{Reasoning quality:} The reasoning quality for both methods was evaluated by a separate instance of the Google Gemini 1.5 Flash LLM, which was prompted to act as an impartial judge. It scored the outputs on a scale where \textbf{1 (Poor)} indicated an irrelevant or ambiguous explanation, and \textbf{2 (Weak)} denoted a plausible but non-specific argument. A score of \textbf{3 (Acceptable)} was given for reasonable but unsupported claims, while \textbf{4 (Good)} was assigned to explanations that used some data but where the analysis remained superficial. The highest score, \textbf{5 (Excellent)}, was reserved for responses that combined multiple, correct, and cited evidence points to construct a firm, evidence-based conclusion.

        \end{enumerate}
\end{itemize}

\section{Results and discussion}
The experimental evaluation focuses on representative Facility Layout Problem (FLP) cases that span variations in facility scale, objective structure, and geometric or operational constraints. These cases capture typical FLP challenges such as minimizing material-handling cost, balancing conflicting objectives, and satisfying constraints like non-overlap, boundary limits, area requirements, and aspect-ratio conditions. Evaluating the proposed method on this diverse set of instances allows us to assess its reasoning capability across problem types commonly encountered in facility-planning contexts. This section interprets the experimental results, compares the performance of the proposed KG-RAG method with the baseline LLM, and discusses the practical implications, failure modes, and operational boundaries revealed through the test cases. Where relevant, the discussion connects these observations to the retrieval components and learning mechanisms described in Section 2 to clarify why specific outcomes were obtained and how they extend to related FLP scenarios.

\begin{figure}[htbp]
    \centering
    \includegraphics[width=1\columnwidth]{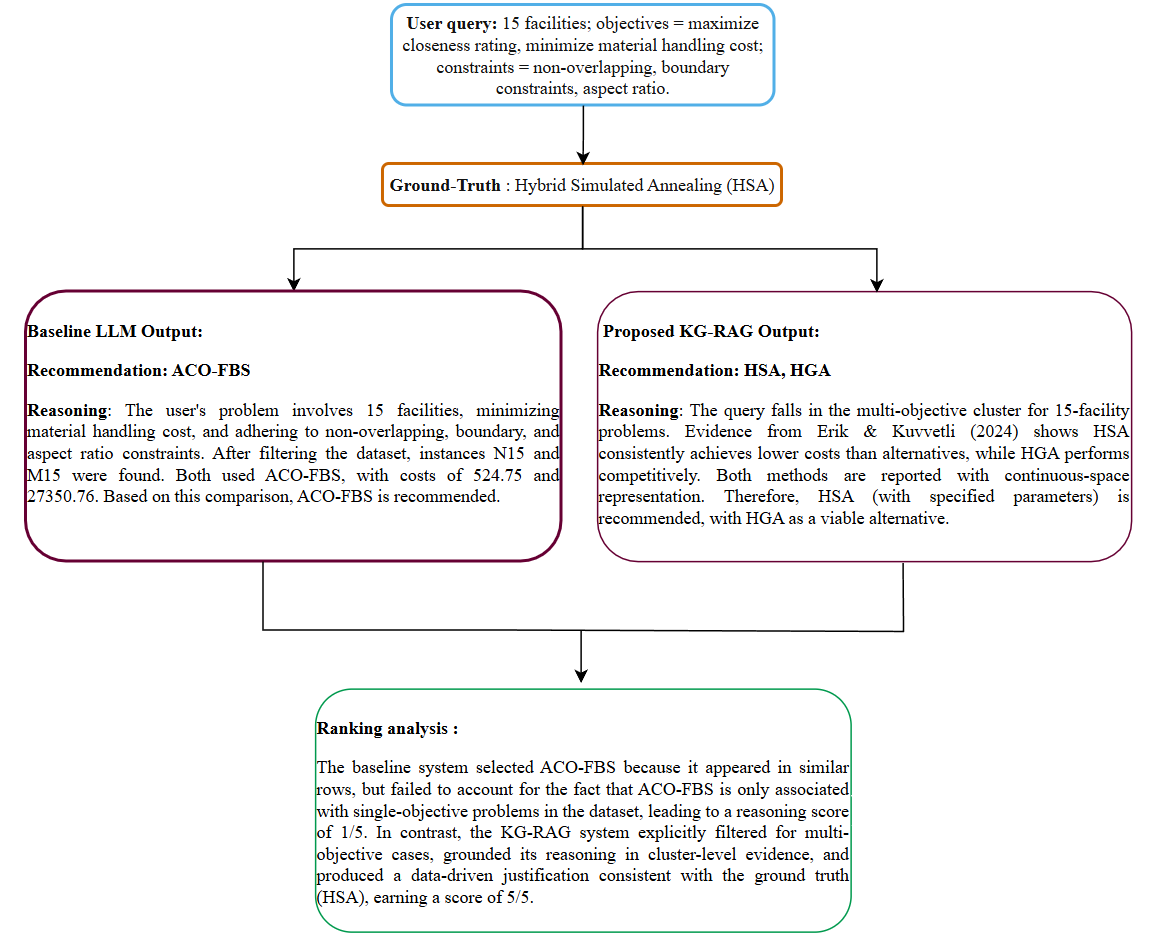}
    \caption{Model recommendation rating analysis }
    \label{fig:rating_analysis}
\end{figure}

The comparative evaluation shows a clear advantage for the KG-RAG method: across the six test cases, it achieved an overall average reasoning score of approximately 4.7/5, while the baseline LLM, when provided with the raw CSV as its knowledge source, achieved an average reasoning score of about 3.3/5, as shown in Table~\ref{tab:comparison} and Figure~\ref{fig:rating_analysis}. These results indicate that the structured relations encoded in the KG materially improve instance-level matching and evidence provenance, whereas the baseline relies primarily on textual co-occurrence and pattern matching. The baseline’s principal error pattern was inferring algorithm suitability from lexical proximity in the CSV rather than from formally encoded relationships. For example, in test case 2 (a multi-objective instance), the baseline recommended ACO-FBS despite the dataset linking ACO-FBS only to single-objective problems. The KG-RAG method avoided this error by restricting retrieval to the multi-objective cluster and by providing the LLM solely with multi-objective precedents. The qualitative reasoning scores further highlight the practical difference between methods. KG-RAG outputs consistently contained traceable citations to problem nodes, method nodes, and cluster trends, which the judging LLM rated as strong evidence-based reasoning. In contrast, baseline outputs often presented plausible but weakly substantiated arguments. Taken together, these findings show that the KG-RAG approach substantially improves explainability and verifiability, even when the evaluation includes boundary-scale cases at the upper limit of KG coverage.
\begin{table*}[htbp]
\caption{Comparative evaluation of recommendation methods}
\label{tab:comparison}
\centering
\begin{tabular}{>{\centering\arraybackslash}p{0.04\linewidth}p{0.35\linewidth}>{\centering\arraybackslash}p{0.1\linewidth}>{\centering\arraybackslash}p{0.1\linewidth}>{\centering\arraybackslash}p{0.07\linewidth}>{\centering\arraybackslash}p{0.13\linewidth}>{\centering\arraybackslash}p{0.07\linewidth}}
\toprule
\textbf{Sl.} & \textbf{Test case} & \textbf{Ground truth} & \textbf{Baseline Rec.} & \textbf{Baseline Rating} & \textbf{KG-RAG Rec.} & \textbf{KG-RAG Rating} \\
\midrule
1 & No of facilities = 10, Objective = min material handling cost, Constraints = non-overlapping, boundary constraints. & CRO-SL & ACO-FBS & 4 & CRO-SL, BRKGA & 5 \\
\addlinespace
2 & No of facilities = 15, Objective = max closeness rating, min material handling cost, Constraints = non-overlapping, boundary constraints, aspect ratio. & HSA & ACO-FBS & 1 & HSA, HGA & 5 \\
\addlinespace
3 & No of facilities = 30, Objective = min material handling cost, Constraints = non-overlapping, boundary constraint, area requirement, aspect ratio. & BRKGA-LP, GA-LP & BRKGA-LP & 5 & BRKGA-LP, GA-LP & 5 \\
\addlinespace
4 & No of facilities = 40, Objective = min material handling cost, Constraints = non-overlapping, boundary constraint, aspect ratio. & BRKGA-LP & ACO-FBS & 3 & BRKGA-LP, ACO-FBS. & 5 \\
\addlinespace
5 & No of facilities = 30, Objective = None, Constraints = None & Construction Heuristic& PROP1 & 4 & BRKGA-LP, Construction Heuristic& 5 \\
 \addlinespace
6 & No of facilities = 125, Objective = None, Constraints = None& BRKGA-LP& SDP& 3& SDP&3\\
\bottomrule
\end{tabular}
\end{table*}

Examining each test case in more depth clarifies how the retrieval channels contributed to the observed outcomes. In the sixth test case (125 facilities), the system did not retrieve the ground-truth BRKGA-LP because the knowledge graph contains very limited evidence at this upper scale, causing both KG-RAG and the baseline to fall back to cluster-level patterns and converge on the same SDP recommendation with a lower reasoning score. In test case 1 (10 facilities, minimize material handling cost), the graph search retrieved several closely related precedents that supported CRO-SL as a proven algorithm for this problem size, while the cluster trends confirmed that BRKGA variants are frequently applied to similar small- and medium-scale instances. The combination allowed the system to recommend both an instance-backed algorithm and a strong alternative. In test case 2 (multi-objective), the restriction of retrieval to the explicitly defined multi-objective cluster prevented a category mismatch and enabled the correct selection of HSA. The baseline failed here by recommending ACO-FBS, which appeared in the dataset under partial feature matches but is only valid for single-objective cases. In test case 4 (40 facilities), exact precedents were absent in the KG. The KG-RAG method addressed this gap by interpolating across a ±25\% facility range and incorporating evidence from the large-scale cluster. This fallback ensured a useful recommendation supported by aggregate patterns, even when instance-level evidence was sparse. Beyond these designed test cases, additional deployment scenarios highlight the broader utility of the framework. In dynamic reconfiguration problems, where objectives change as layouts evolve, the pipeline can requery objective and scale clusters to surface algorithms historically applied under similar shifts, providing actionable guidance without requiring new data collection. For highly constrained layouts involving adjacency or aspect-ratio conditions, the graph’s explicit links between problems and constraint-handling strategies make it possible to identify hybrid algorithms, such as metaheuristics combined with LP repair operators, that have succeeded in comparable contexts. Finally, for novice users or under-specified queries, the combined use of cluster-level trends and vector similarity led the system to propose both a conservative starting point (a construction heuristic) and a higher-performance alternative (a BRKGA variant), demonstrating how the retrieval channels interact to generate useful recommendations even when minimal information is available.

While the KG-RAG method consistently outperformed the baseline, the experiments also revealed several concrete limitations and failure modes relevant to practical use. The most direct limitation is the dependency on KG coverage. When the KG lacks examples for a specific region of the problem space, such as uncommon facility scales or rare combinations of constraints, the precision of exact-match retrieval decreases, and the system relies more heavily on cluster-based fallbacks. This behaviour was clearly reflected in the sixth test case (125 facilities), where sparse upper-scale coverage caused both methods to depend on coarse cluster-level evidence, resulting in a non-ground-truth recommendation with a lower reasoning score. This reliance ensures that recommendations remain available, but it can reduce specificity at the algorithm–instance level. A second limitation arises from the ingestion process. If user-submitted cases are incorrectly formatted or contain low-quality information, they may introduce noisy or misleading links into the KG, which in turn can affect retrieval accuracy. Third, although grounding substantially reduces the risk of hallucination, it does not fully eliminate overgeneralization. When the assembled dossier for the LLM contains sparse or weak evidence, the model may still generate plausible but insufficiently supported explanations. Fourth, scaling to very large deployments increases computational demands: repeated embedding generation and exhaustive similarity searches can introduce latency and backend load that affect real-time responsiveness. These patterns were evident in the evaluation whenever exact algorithm precedents were missing. In such cases, the system produced cluster-level recommendations that were logically sound but lacked the precision of instance-level matches. Another boundary condition appears in out-of-scope queries, where the problem specifications fall far outside the range of KG data. Rather than failing outright, the system defaults to the most relevant high-level clusters, providing a general but still actionable recommendation. This behaviour illustrates both the resilience and the boundaries of the framework: it can continue to return useful outputs, but high-precision recommendations require representative KG coverage, careful ingestion, and efficient retrieval management.

To address the observed limitations without changing the overall KG-RAG design, several practical measures can be applied. During ingestion, attaching provenance metadata and simple quality scores allows the retrieval engine to weight evidence by reliability, while a lightweight human-in-the-loop review prevents long-term degradation of the KG. Embeddings can be improved by adding structured domain features, such as facility counts or aspect ratios, alongside text descriptions to enhance vector search fidelity. Prompt templates and dossier rules should enforce a minimum evidence threshold before returning strong claims; when this is not met, the system can provide lower-confidence recommendations and suggest data enrichment. Operationally, incremental embedding updates, approximate nearest-neighbour indices, and caching reduce latency as the graph grows. Further analyses would clarify robustness and the contribution of individual components. Ablation studies disabling each retrieval channel (graph-only, vector-only, or cluster-only) would quantify their marginal benefits. Sensitivity analysis masking parts of the KG would reveal how coverage affects performance and guide data acquisition. User studies with domain experts would assess trust and usability, while latency profiling across KG sizes would inform engineering trade-offs. The results also indicate conditions that affect deployment. Performance is strongest when the KG contains diverse, well-labelled precedents covering the target problem range; the behaviour observed in the sixth test case, where sparse upper-scale coverage reduced retrieval precision, highlights the importance of continued data growth. When exact precedents are missing, the fallback to cluster-level evidence produces consistent but less specific recommendations; presenting these with confidence indicators helps practitioners interpret them appropriately. The explicit evidence trail ensures recommendations remain transparent and auditable, supporting the system’s role as a decision-support tool.

\section{Conclusion}
\label{sec:conclusion}
Finally, this section concludes by summarizing the primary contributions and key findings of the paper, discussing the current limitations of the method and highlighting promising directions for future research. This paper introduced a new recommendation method for the FLP based on a KG-RAG framework. The method integrates a domain-specific KG with a multi-stage retrieval pipeline and an LLM to deliver explainable, data-driven algorithm recommendations. In evaluation, the proposed method consistently outperformed the baseline Gemini 1.5 Flash LLM provided with the same knowledge base in tabular form, achieving an overall average reasoning score of approximately 4.7 across six test cases, compared to about 3.3 for the baseline. These results highlight the strength of combining structured knowledge with hybrid retrieval strategies, enabling the system to generate context-aware and explainable algorithm recommendations, even in complex scenarios where simpler baseline approaches may struggle. The primary contribution of this work is the design of a multi-stage KG-RAG pipeline that combines graph-based, vector-based, and cluster-based search to assemble context-specific evidence. A key distinction from the baseline is the use of a structured KG rather than a flat CSV table, which enables reasoning over formally encoded relationships instead of relying only on superficial matches. This design leads directly to two major benefits: first, the ability to generate explainable recommendations supported by traceable evidence from problem nodes, method nodes, and cluster trends; and second, a substantial reduction in unsupported or hallucinated outputs, as the LLM is grounded in structured domain knowledge. Together, these contributions demonstrate how the proposed approach advances beyond standard LLM-based recommendations and provides a reliable decision-support framework for FLP.

The study also revealed boundaries that define the current scope of the method. Performance depends on the coverage and quality of the underlying KG, and the sixth test case, which occurs near the upper limit of the current KG illustrated how sparse coverage can reduce retrieval precision and lead to broader, cluster-level recommendations. Recommendations in out-of-scope queries are therefore necessarily based on categorical evidence rather than precise precedents. Ingestion of noisy or low-quality cases may also affect retrieval fidelity, and scaling to very large KG introduces computational demands. These limitations are not fundamental flaws but natural boundaries, and they can be progressively addressed through provenance-aware ingestion, efficient retrieval mechanisms, and continuous graph expansion via the feedback loop. Looking ahead, future work will extend the framework to other combinatorial optimization problems, evaluate scalability across larger and more diverse datasets, and further strengthen continuous update mechanisms. By combining structured knowledge with hybrid retrieval and language model reasoning, the KG-RAG framework establishes a robust foundation for explainable and data-driven algorithm recommendations in complex optimization domains. 

\section*{Acknowledgement}
The authors express their sincere gratitude to TCS Research, Tata Consultancy Services Ltd., especially to the CTO team, for providing the internship opportunity and valuable guidance that greatly contributed to the completion of this work.


\end{document}